\documentclass[pra,showpacs,amsmath,twocolumn]{revtex4-1}
\usepackage[colorlinks=true,urlcolor=blue,citecolor=blue,linkcolor=blue]{hyperref}
\usepackage[sort&compress]{natbib}
\setcitestyle{numbers,square}
\usepackage[english]{babel}
\usepackage{setspace}
\usepackage{amssymb}
\usepackage{amsthm}
\usepackage{amscd}
\usepackage{rotating}
\usepackage{color}
\usepackage{graphics, graphicx}
\usepackage{psfrag}
\usepackage{mathcomp}
\usepackage{accents}

\newcommand*{\dt}[1]{%
  \accentset{\mbox{\large\bfseries .}}{#1}}

\begin{document}
\title{Dynamics of Disordered States in the Bose-Hubbard Model with Confinement}
\author{Mi Yan, Hoi-Yin Hui, and V. W. Scarola}
\affiliation{Department of Physics, Virginia Tech, Blacksburg, Virginia 24061}

\begin{abstract}
Observations of center of mass dynamics offer a straightforward method to identify strongly interacting quantum phases of atoms placed in optical lattices.  We theoretically study the dynamics of states derived from the disordered Bose-Hubbard model in a trapping potential.  We find that the edge states in the trap allow center of mass motion even with insulating states in the center.  We identify short and long-time scale mechanisms for edge state transport in insulating phases.  We also argue that the center of mass velocity can aid in identifying a Bose-glass phase.  Our zero temperature results offer important insights into mechanisms of transport of atoms in trapped optical lattices while putting bounds on center of mass dynamics expected at non-zero temperature.
\end{abstract}
\date{\today}

\maketitle

\section{Introduction}

Ultracold atomic and molecular gases placed in optical lattices realize strongly correlated states of matter captured by Hubbard models \cite{jaksch:1998,greiner:2002,lewenstein:2007,bloch:2008}.  Hubbard models are known to support fundamentally important states that serve as paradigms for understanding quantum collective behavior in solids.  New methods to observe and study these collective phases in optical lattices have been invented to better probe interesting quantum behavior.  Example diagnostics include Bragg spectroscopy \cite{birkl:1995,weidemuler:1995} (applied recently to measure \cite{hart:2015} local  antiferromagnetic spin correlations \cite{corcovilos:2010,fuchs:2011,kozik:2013}), noise spectroscopy \cite{altman:2004,foelling:2005,spielman:2007}, and center of mass transport.

Center of mass transport experiments in optical lattices are often used to probe response in direct analogy to transport in solids.  In optical lattice transport the center of mass is shifted and the dynamics is observed using time of flight.  Transport studies of bosonic atoms in optical lattices have been used to study superfluid and Mott insulating states, states of the Bose-Hubbard model \cite{cataliotti:2001,cristiani:2004,fallani:2004,fertig:2005,desarlo:2005,henderson:2006,mun:2007,fallani:2007,mckay:2008,white:2009,pasienski:2010}.  When the system is in a superfluid state it oscillates within the trapping potential.  The Mott insulator shows, by contrast, essentially no dynamics for very strong interactions.  More recently, disorder has been implemented \cite{lye:2005,clement:2005,FortFallani2005,lye:2007,fallani:2007,billy:2008,roati:2008,chabe:2008,chen:2008,Clement2008,aspect:2009,Fallani2008119,sanchez-palencia:2010,white:2009,pasienski:2010,robert:2010,tanzi:2013,Modugno2010} to further probe glassy states expected from a competition between strong disorder and inter-particle interactions.  In the experiments of Ref.~\cite{pasienski:2010} it was argued that the suppression of center of mass transport indicated the presence of a Bose-glass, a state of the Bose-Hubbard phase diagram induced by disorder.

\begin{figure}[t]
  \includegraphics[width=\linewidth]{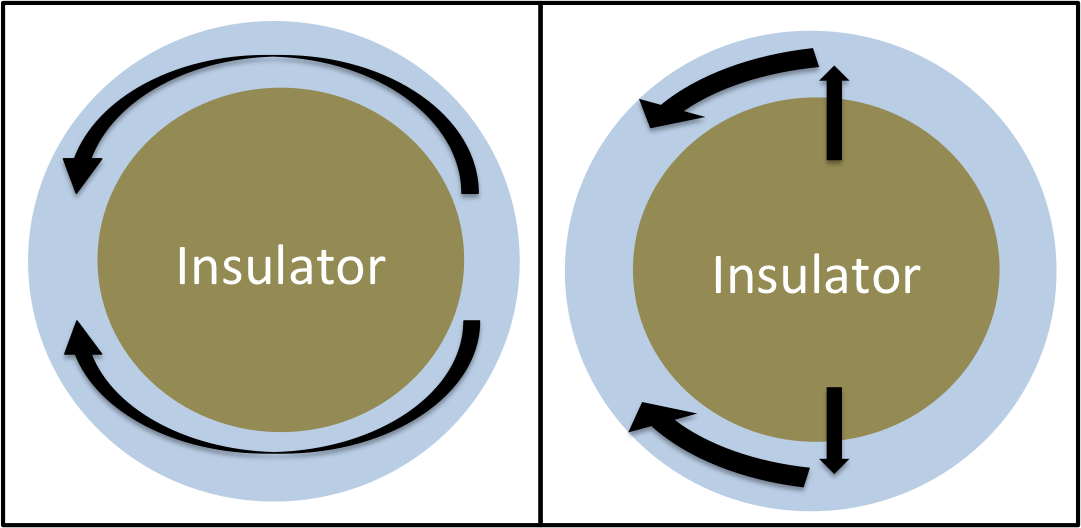}
  \caption{(Color online) Schematic showing two distinct mechanisms for center of mass dynamics after an initial shift of the trapping potential to the left.  The center of the system is in an insulating phase (either the Mott insulator or the Bose-glass) but the edges are superfluid. Left Panel:  We find that initially (on times scales shorter than the inter-site tunneling) the edge superfluid quickly moves around the insulator to the left to redistribute the total center of mass.  Right Panel:  We find that (on time scales much larger than the inter-site tunneling) a two stage process shifts the center of mass.  First particles slowly tunnel out of the central insulator into the edges (vertical arrows).  Then the particles quickly move to the left along the edge to redistribute the center of mass.}
  \label{fig_schematic_flow}
\end{figure}

Optical lattice transport experiments have fundamental and crucial differences from solid state transport.  Particles in optical lattices are, to a good approximation, isolated systems.  The absence of a heat bath prevents direct thermalization.  Thermalization of initial states occurs via inter-particle interactions \cite{mckay:2011}.  The system can therefore be slow to thermalize during transport because there is no external bath.  Furthermore, the absence of lattice phonons avoids conventional assumptions of dephasing often invoked in solids.  Other thermal effects are also very different.  For example, the absence of a particle number bath prevents transport via variable range hopping.  But the most pronounced differences arise from strong trapping.  In optical lattice experiments the response of the center of mass must be understood in the context of a parabolic trapping potential which mixes phases in the trap and leads to edge states \cite{jaksch:1998,lewenstein:2007,scarola:2007}.  

In this work we study the dynamics of bosons in a disordered Bose-Hubbard model in the presence of a trap.  We seek to identify the precise mechanisms of transport in a trap and correlate the response with known phases.  We will include disorder to study the dynamic response of the following phases to an external field: a disordered superfluid, a Bose-glass, and a disordered Mott insulator (as in Ref.~\cite{lin:2012} but in the presence of confinement).  We will also focus entirely on the zero temperature response to identify purely quantum transport effects that should underpin low temperature experiments.  

We find that the insulating states (Bose-glass and Mott insulator) have center of mass motion that is dominated by edge effects.  Our simulations show that when the center of the system is in an insulating phase, it remains motionless under a trap shift, while the edge states move, and therefore, so does the center of mass.  Fig.~\ref{fig_schematic_flow} shows a schematic depicting distinct mechanisms of transport that we find.  The left panel depicts what we find at short times (with respect to the single-particle inter-site tunneling time scale).  Here we find that the edge states surrounding an insulator quickly move and dominate the center of mass motion.  

But at long times we find that disordered insulating states drift slowly (the center of mass moves only a site or two for times on the order of 100 times the inter-site tunneling).  The insulating states reach a terminal velocity that does not change with the applied field.  This is contradictory to the conventional Drude-type picture of particle motion.  We instead propose a two-stage process in which slow tunneling into edge states (right panel of Fig.~\ref{fig_schematic_flow}) slowly moves the center of mass while leaving the central insulator motionless.  

Our results reveal mechanisms for center of mass transport.  The presence of a trap leads to edge states.  The edge states are responsible for motion of the center of mass via two distinct mechanisms, one at short times and one at long times.  These mechanisms also differ from conventional pictures of bulk transport in solids where an applied field leads to an average drift of all particles.  Our results establish a basis for interpreting measurements of center of mass dynamics of atoms trapped in optical lattices.

\section{Model}
\label{sec_model}

Ongoing optical lattice experiments containing bosonic alkali atoms are accurately captured by the Bose-Hubbard model \cite{jaksch:1998}.  Controlled disorder, applied through speckle laser light or incommensuration \cite{lewenstein:2007}, can significantly alter the phase diagram and therefore impact transport.  Transport experiments \cite{pasienski:2010} can be understood from an interplay of temperature, disorder, trapping, and interactions \cite{scarola:2015}. In this work we study zero temperature effects to focus on just disorder, trapping, and interactions.   

We study the dynamics of states of the disordered Bose-Hubbard model to assess transport properties in a parabolic trapping potential.  The Bose-Hubbard Hamiltonian is given by:
\begin{eqnarray}
H = - t \sum_{\langle i,j \rangle}  b_i^\dagger b_j  + \frac{U}{2}\sum_i \hat n_i (\hat n_i - 1) +\sum_i  \mu_i \hat n_i,
 \label{bhmodel}
\end{eqnarray}
where $b_{i}^{\dagger}$ creates a boson at the site at lattice position ${\mathbf{R}_i}$ and $\hat{n}_{i}= b_{i}^{\dagger} b_{i}^{\phantom\dagger}$ is the number operator.  Here $\langle i,j\rangle$ denotes summation over nearest neighbor sites.  
$t$ is the single-particle nearest neighbor tunneling amplitude and $U$ is the on-site repulsive interaction energy.  The spatially inhomogeneous  chemical potential is: 
\begin{eqnarray}
\mu _{i}=-\mu+ \Omega \vert {\mathbf R}_i - {\mathbf R}_0(\tau) \vert^2+\epsilon_{i},
\label{eq_muj}
\end{eqnarray}
where the central chemical potential, $\mu$, tunes the average density. $\Omega$ is the strength of the harmonic trapping potential and $\vert {\mathbf R}_i - {\mathbf R}_0(\tau) \vert$ defines the distance between a site $i$ and the trap center, ${\mathbf R}_0(\tau)$.  We work on a two-dimensional square of side lengths $L=L_{x}=L_{y}$.  We choose $L$ so that the density for edge sites vanishes for all trap frequencies used.  Distances are measured in units of the lattice spacing, $a$.  The trap center is a function of time, $\tau$, and, in what follows, will be immediately shifted along the negative $x$ direction a distance $\Delta R_{0}$ to induce particle number flow at $\tau>0$.  The total pulse sequence is then: $ {\mathbf R}_0(\tau)=0$ for $\tau\leq 0$ and ${\mathbf R}_0(\tau)= -|\Delta R_0| \hat{x}$ for $\tau> 0$.  By expanding the squared term in $\mu_{i}$ we see that the trap shift can be thought of as an applied linear potential: $\sim 2\Omega {\mathbf R}_i \cdot \hat{x} (\Delta R_0)$.

\begin{figure}[t]
  \includegraphics[width=0.9\linewidth]{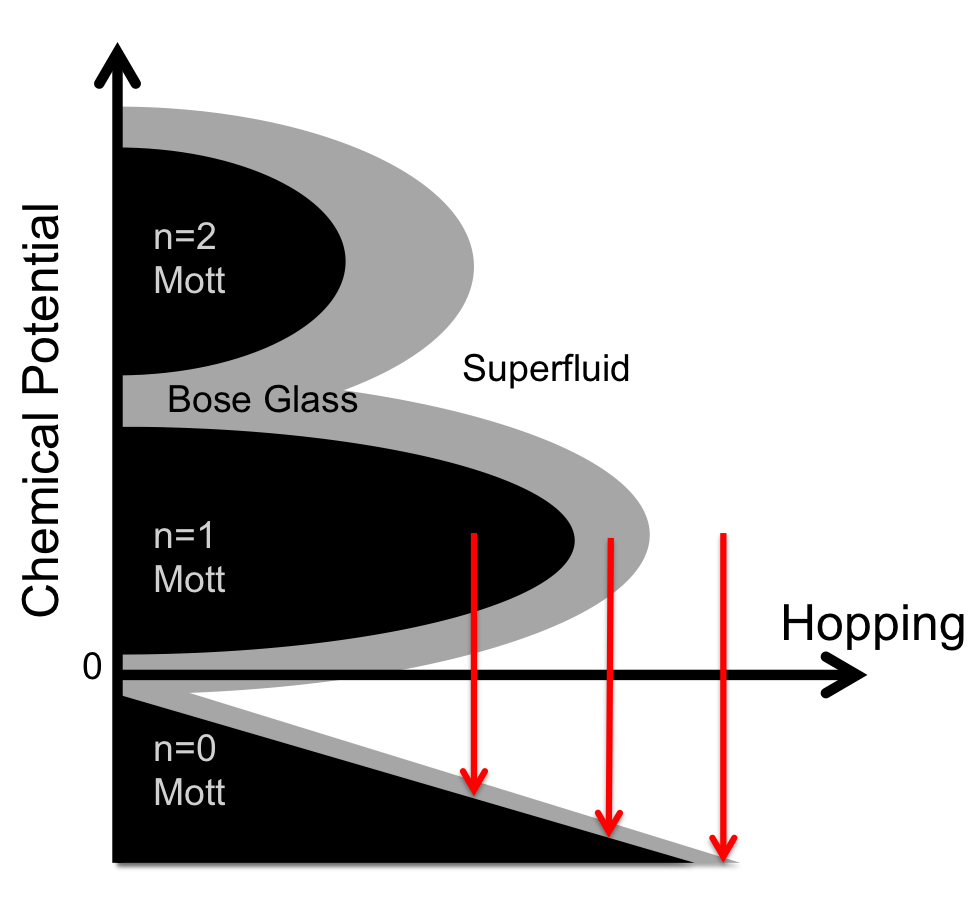}
  \caption{ (Color Online) Schematic mean field phase diagram of the disordered Bose-Hubbard model [Eq.~(\ref{bhmodel})] in the absence of a trap, $\Omega=0$. The black regions indicate the incompressible Mott insulators found at integer densities.  The grey regions indicate the compressible Bose-glass that separates the Mott from the superfluid.  The three vertical arrows depict how the trapping potential lowers the local chemical potential as we move from the center of the system (top of the arrow) to the edge (bottom of the arrow).  The three arrows represent regimes studied here, where the center of the system contains a:  disordered Mott insulator (left), Bose-glass (center), and disordered superfluid (right).  }
  \label{fig_phasediagram}
\end{figure}

$\epsilon_i$ denotes a random energy shift at the $i$th site.   In our study, the disorder is uniformly distributed with $\epsilon_i\in\{-\Delta,\Delta\}$.  Specifically, we use the box probability distribution function:
 \begin{eqnarray}
  P(\epsilon)=\frac{\Theta \left(\Delta-\vert\epsilon\vert \right)}{2\Delta},
   \label{eq_PU}
 \end{eqnarray}
where $\Delta$ is the strength of the disorder and $\Theta$ is the Heaviside step function. 

\begin{figure}[t]
\centering
  \begin{tabular}{@{}cc@{}}
    \includegraphics[width=.16\textwidth]{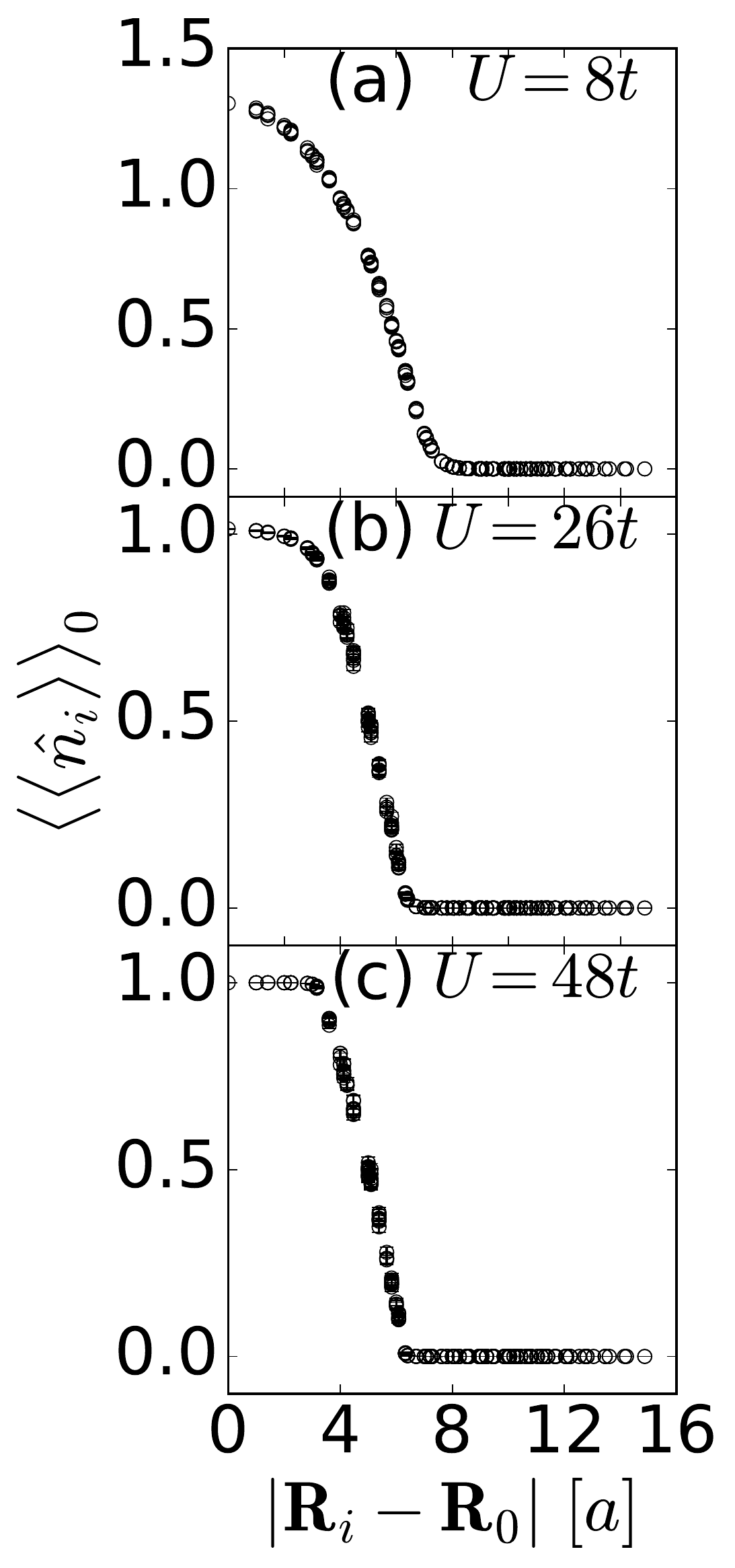} &
    \includegraphics[width=.17\textwidth]{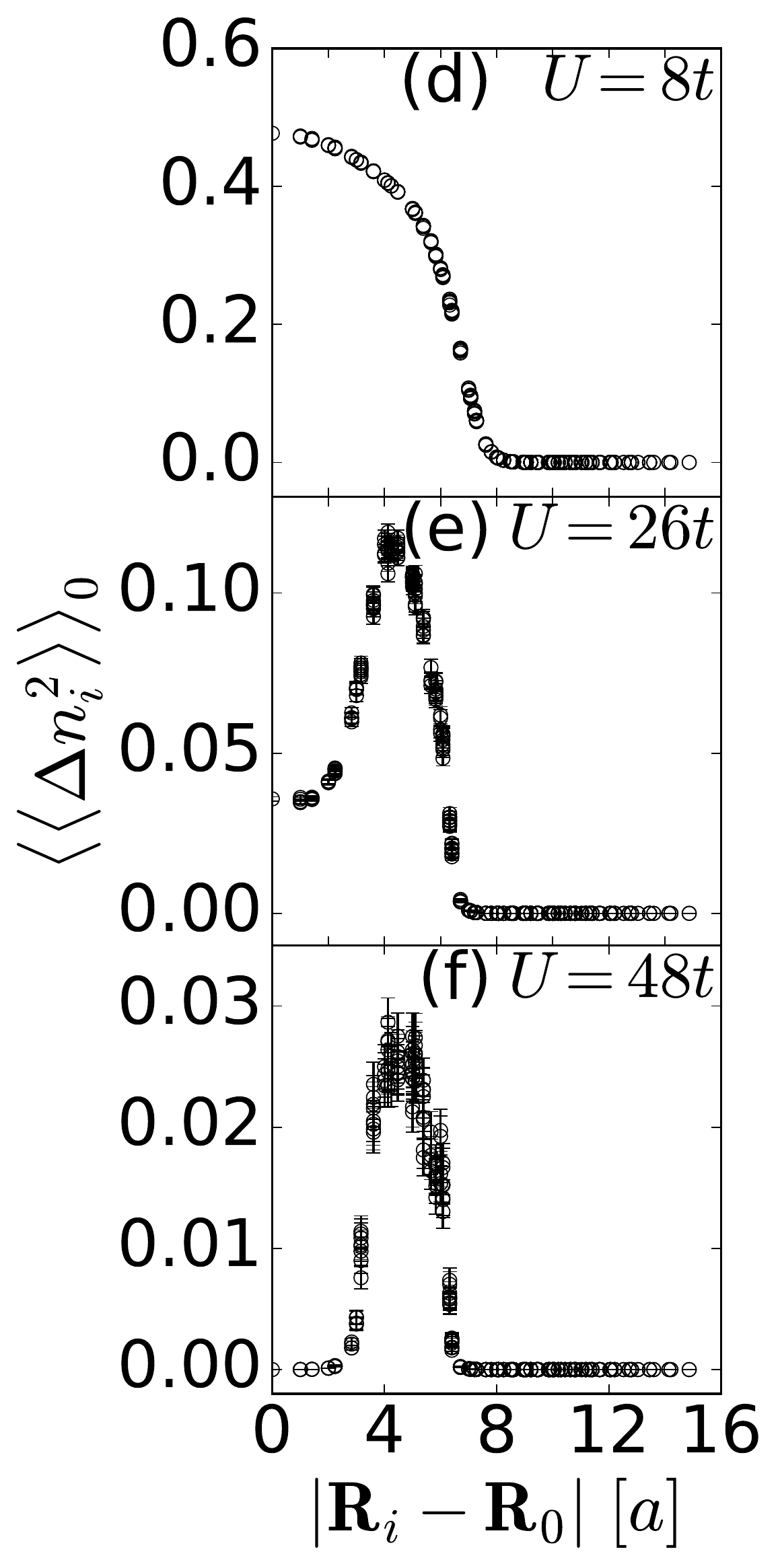} 
        \includegraphics[width=.16\textwidth]{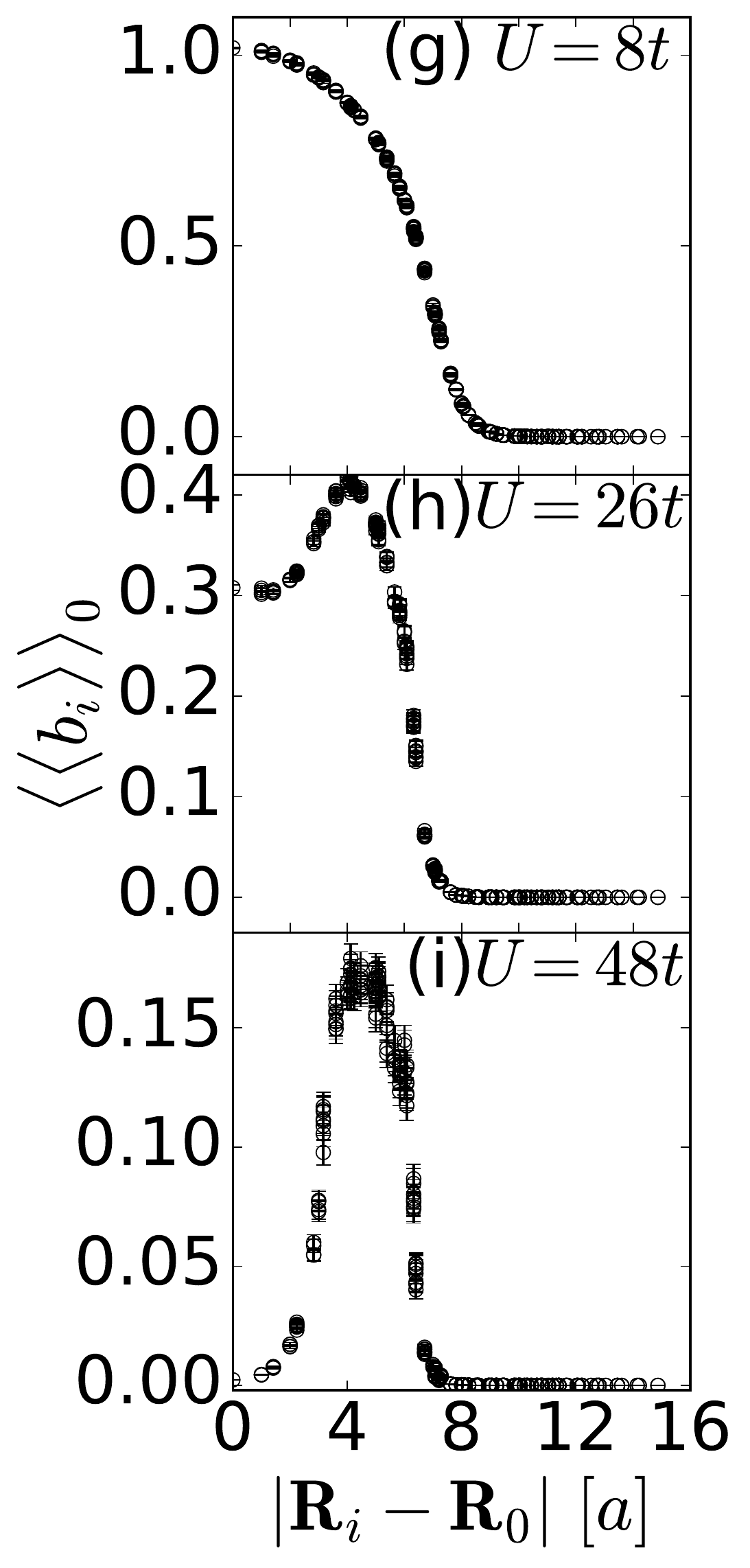} 
        
\end{tabular}
  \caption{Panels (a)-(c) plot the disorder averaged local density in the initial state as a function of distance from the center of the trap.  Panels (a) and (c) have a superfluid and Mott insulator in the trap center, respectively.  Panel (b) is in an intermediate regime.  Model parameters are chosen to be: disorder strength $\Delta=0.3U$, trap strength $\Omega=0.02U$, and central chemical potential $\mu=0.5U$.  Panels (d)-(f) plot the same but for the disorder averaged local density fluctuations in the initial state.  The density fluctuations in panels (e) and (f) show the edge superfluid. Panels (g)-(i) plot the disorder averaged local superfluid order parameter in the initial state. Error bars result from disorder averaging.}
  \label{fig_density_radius}
\end{figure}

\begin{figure}[t]
\centering
  \begin{tabular}{@{}cc@{}}
    \includegraphics[width=.24\textwidth]{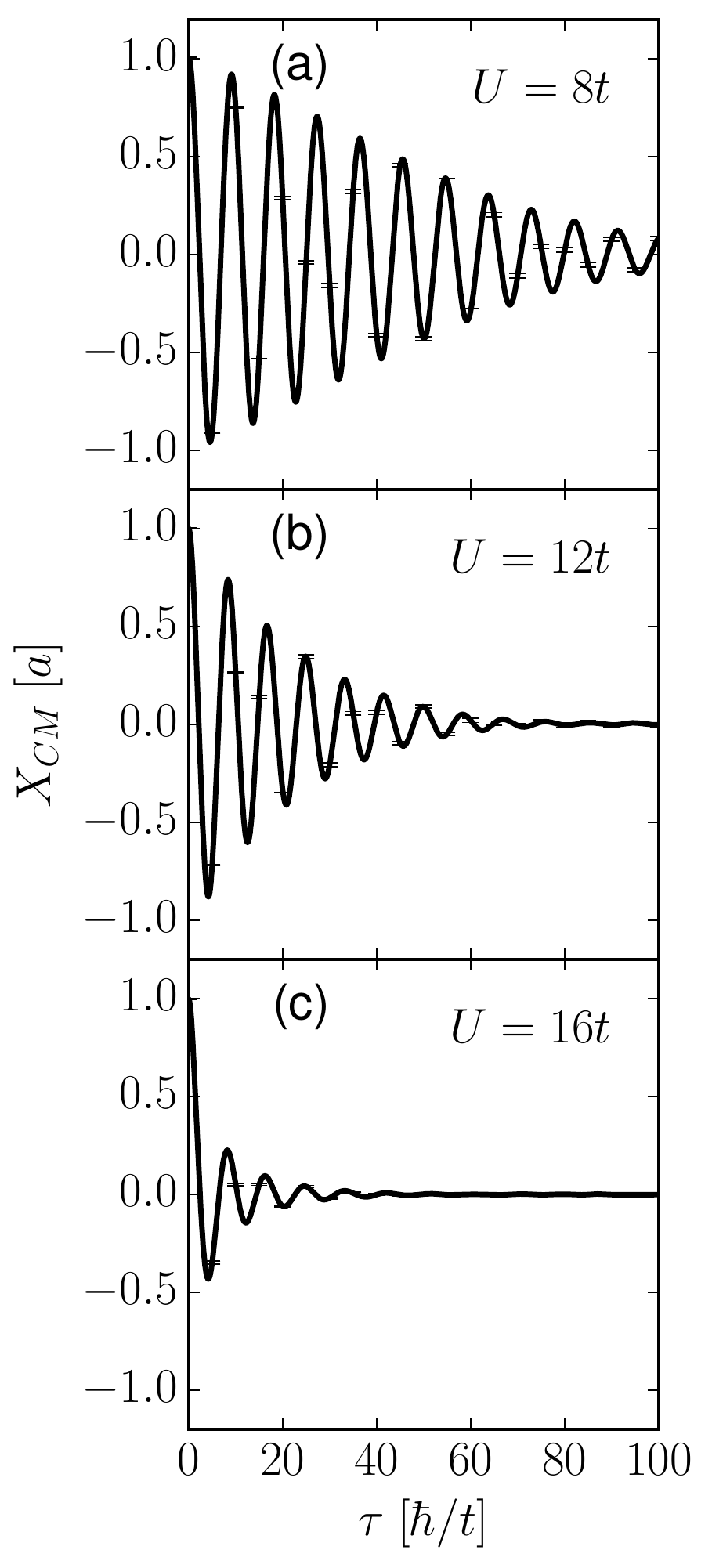} &
    \includegraphics[width=.23\textwidth]{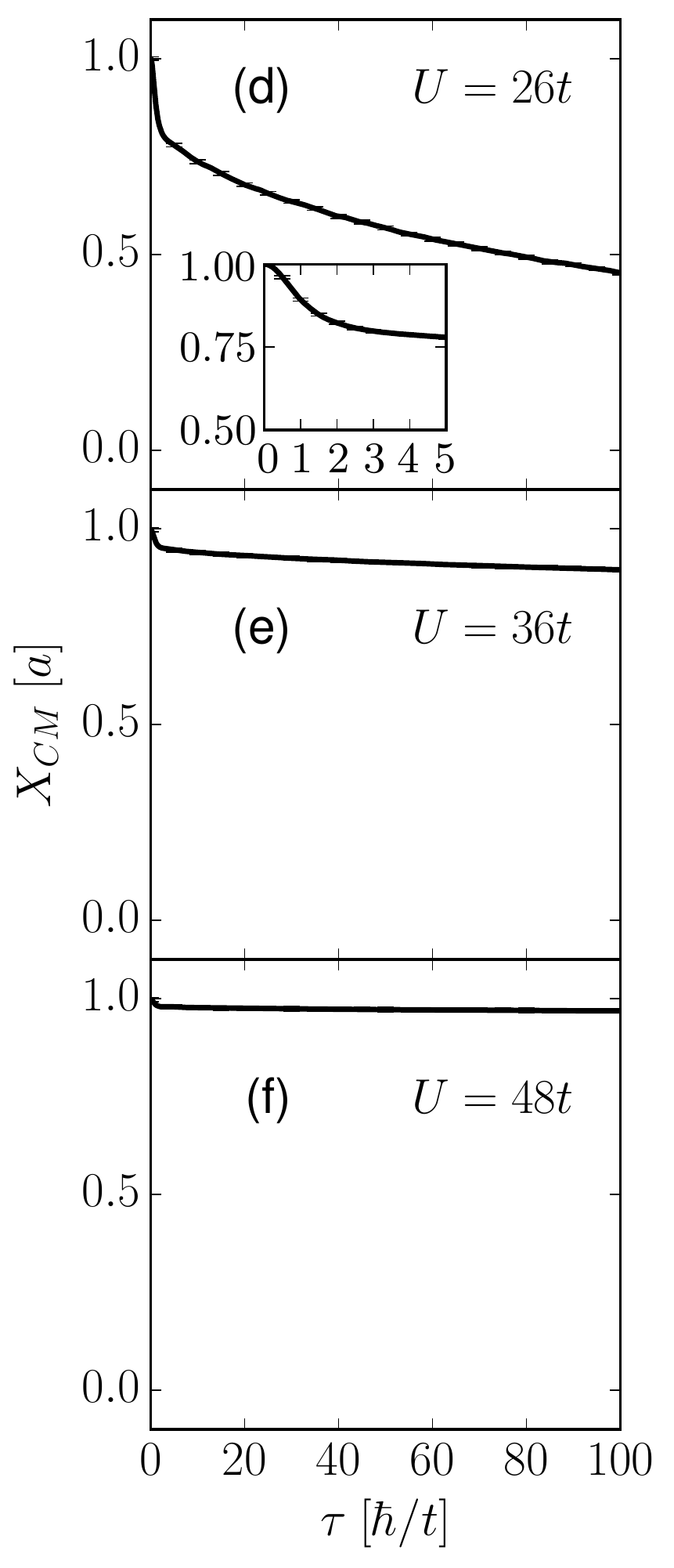} 
\end{tabular}
  \caption{The disorder averaged center of mass position along the $x$-direction plotted as function of time for various interaction strengths, $U$.  The trap shift is chosen to be $\Delta R_{0}=a$ and other model parameters are the same as Fig.~\ref{fig_density_radius}.  Panels (a)-(c) are initially in the disordered superfluid state. In panel (d) the center of the system is initially in the Bose-glass state whereas panel (f) is initially in the Mott state.  The inset to panel (d) shows the short-time behavior. Panel (e) is initially in an intermediate state.  In panels (d)-(f), the center of mass has a nearly constant velocity for $\tau\gg\hbar/t$. }
  \label{fig_com_time_8_48}
\end{figure}

To study the dynamics we use a stochastic mean field decoupling performed using variational wavefunctions: the time-dependent Gutzwiller ansatz \cite{rokhsar:1991,jaksch:2002, JreissatyCarrasquillaWolfEtAl2011,*JreissatyCarrasquillaRigol2013}.  The ansatz yields wavefunctions whose static expectation values agree very well with quantum Monte Carlo away from phase boundaries in 2D in 3D (see, e.g., Refs.\cite{zakrzewski:2005} and \cite{hen:2010a}).  The wavefunction at time $\tau$ is assumed to be in the product form:
\begin{equation}
|\Psi_{\rm GW}(\tau)\rangle=\prod_{i}\left[\sum_{n=0}^{\infty}f_n^{(i)}(\tau) |n\rangle_i \right],
\label{eq_wf}
\end{equation}  
where $f^{(i)}_{n}(\tau)$ is the complex amplitude to find $n$ atoms on lattice site $i$ at time $\tau$.  

The initial state can be found using the variational theorem.  We first assume a random initial guess for all of the $f$'s.  We then minimize the energy of $H$ by varying all $f$'s using a conjugate gradient method.  We find that keeping the parameters $f_{n}$  for $n\leq 4$ is sufficient for the regime studied here.  Once the wavefunction is found, all initial state correlation functions can be computed.

\begin{figure*}[t]
  \includegraphics[width=\textwidth]{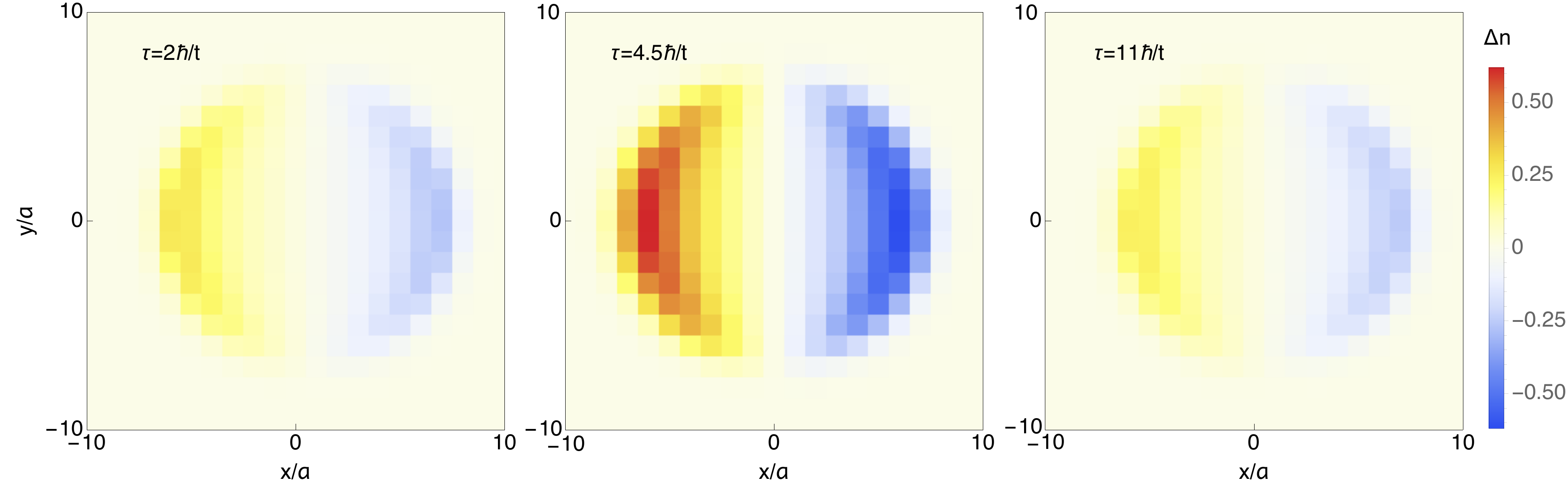}
  \caption{(Color online) The disorder averaged local density deviation from the initial state, $\Delta n_{i}\equiv\langle \langle n_{i} \rangle \rangle_{\tau} -\langle \langle n_{i} \rangle \rangle_{\tau=0}$, plotted as a function of position in the $x-y$ plane for three different times.  All model parameters are the same as Fig.~\ref{fig_com_time_8_48}a where a superfluid occupies nearly the entire trap.  These parameters correspond to the right arrow in the phase diagram, Fig.~\ref{fig_phasediagram}.  Here we see how the superfluid as a whole oscillates in the trap even in the presence of disorder.  }
  \label{fig_density_U_8}
\end{figure*}

We propagate the initial state wavefunction in time using the equations of motion consistent with the Gutzwiller ansatz.  The equations of motion for $f^{(i)}_{n}$ are \cite{jaksch:2002}:   
\begin{eqnarray}
\mathrm{i\hbar}\partial_{\tau}{f}^{(i)}_n &=& \left[\frac{U}{2}n(n-1)+n \mu_i\right]f^{(i)}_n\\ \nonumber
&-&t\sum_{\langle i, j \rangle}\left(\psi^{*}_j \sqrt{n+1} f^{(i)}_{n+1}+ \psi_{j} \sqrt{n} f^{(i)}_{n-1}\right), 
\end{eqnarray} 
where $\psi_{j} = \langle b_j \rangle = \sum_{n}\sqrt{n}f^{(j) *}_{n-1}f^{(j)}_{n} $ is the mean-field superfluid order parameter.
We numerically solve this set of coupled first order differential equations using the adaptive step size Runge-Kutta method.  This allows access to all correlation functions as a function of time.

The center of mass position and velocity are key observables often used in experiments to reveal insulating behavior.  We compute the total center of mass position along the $x$-direction and the center of mass velocity as a function of time:
\begin{eqnarray}
X_{\text{CM}}(\tau)&=&\frac{1}{N}\sum_{i} \langle \langle \hat{n}_{i} \rangle \rangle_{\tau} [\hat{x}\cdot {\mathbf R}_i] \nonumber\\
V_{\text{CM}} (\tau) &=& \dt{X}_{\text{CM}}(\tau).
\end{eqnarray} 
Here $N$ is the number of particles and $\langle \langle ... \rangle \rangle _{\tau}$ indicates the disorder average of quantum state averages with respect to $|\Psi_{\rm GW}(\tau)\rangle$.  We find that disorder averaging with 1000 configurations is sufficient to obtain convergence.  

\section{Static Properties}
\label{sec_statics}

In this section we review the static properties of the disordered Bose-Hubbard model at zero temperature.  The schematic mean field phase diagram \cite{fisher:1989} reproduced by the Gutzwiller ansatz \cite{buonsante:2009} is depicted in Fig.~\ref{fig_phasediagram}.  Here we see that the Mott insulator is separated from the superfluid by an intermediate Bose-glass.  The Mott insulator occurs at integer filling and is incompressible.  It has zero condensate fraction. The Bose-glass is compressible but has non-zero local condensate fraction due to superfluid puddles embedded in an insulating background. The superfluid is characterized by a non-zero stiffness \cite{fisher:1973} and compressibility.

The trapping potential lowers the chemical potential in moving from the center to the edge of the trap.  The trap can (at a mean field level) be thought of as sweeping the chemical potential vertically through the phase diagram.  The vertical arrows in Fig.~\ref{fig_phasediagram} depict the three regimes of the trapped disordered Bose-Hubbard model studied in this work.  The left, central, and right arrows indicate parameters such that the central region of the trap (the top of the arrow) hosts a disordered Mott insulator, Bose-glass, and disordered superfluid, respectively.  Due to the inhomogeneous trapping potential, superfluid and Bose-glass coexist in the edge of the trap in all three regimes.  

To characterize the phases in the trap we compute local correlation functions.  Figs.~\ref{fig_density_radius}a-c plot the local density as a function of distance from the trap center.  The density qualitatively reveals the location on the phase diagram, with $\langle \hat{n} \rangle\approx1$ indicating the Mott insulator.  The local density fluctuations:
\begin{eqnarray}
\langle \Delta n_{i}^{2} \rangle \equiv \langle \hat{n}_{i}^{2} \rangle - \langle \hat{n}_{i} \rangle^{2},
\end{eqnarray} 
capture the local compressibility since: $\langle \Delta n_{i}^{2} \rangle \propto \kappa_{i} k_{B}T$, where $T$ is temperature and local compressibility is $\kappa_{i}=\partial \langle \hat{n}_{i} \rangle/\partial \mu$.  Figs.~\ref{fig_density_radius}d-f plot the density fluctuations as a function of distance from the trap center. The sharp increase of the fluctuations at the edge (in Figs.~\ref{fig_density_radius}e and \ref{fig_density_radius}f) indicate compressible edge states.  We have also used the local superfluid order parameter (in Figs.~\ref{fig_density_radius}g-i) and other correlation functions to identify the location on the phase diagram.

\section{Center of Mass Dynamics}
\label{sec_shortcm}

This section presents our primary results by correlating center of mass dynamics with phases in the center of the trap.  We find that even with moderately strong disorder, the superfluid phase oscillates.  Strong interaction significantly damps superfluid motion.  But when the center of the trap enters the Bose-glass and Mott phases, the center of mass dynamics slows considerably.  By computing the density in the lattice we find that the center of the system fails to move. Here the dynamics is dominated by flow along the edges.

\begin{figure*}[t]
  \includegraphics[width=\textwidth]{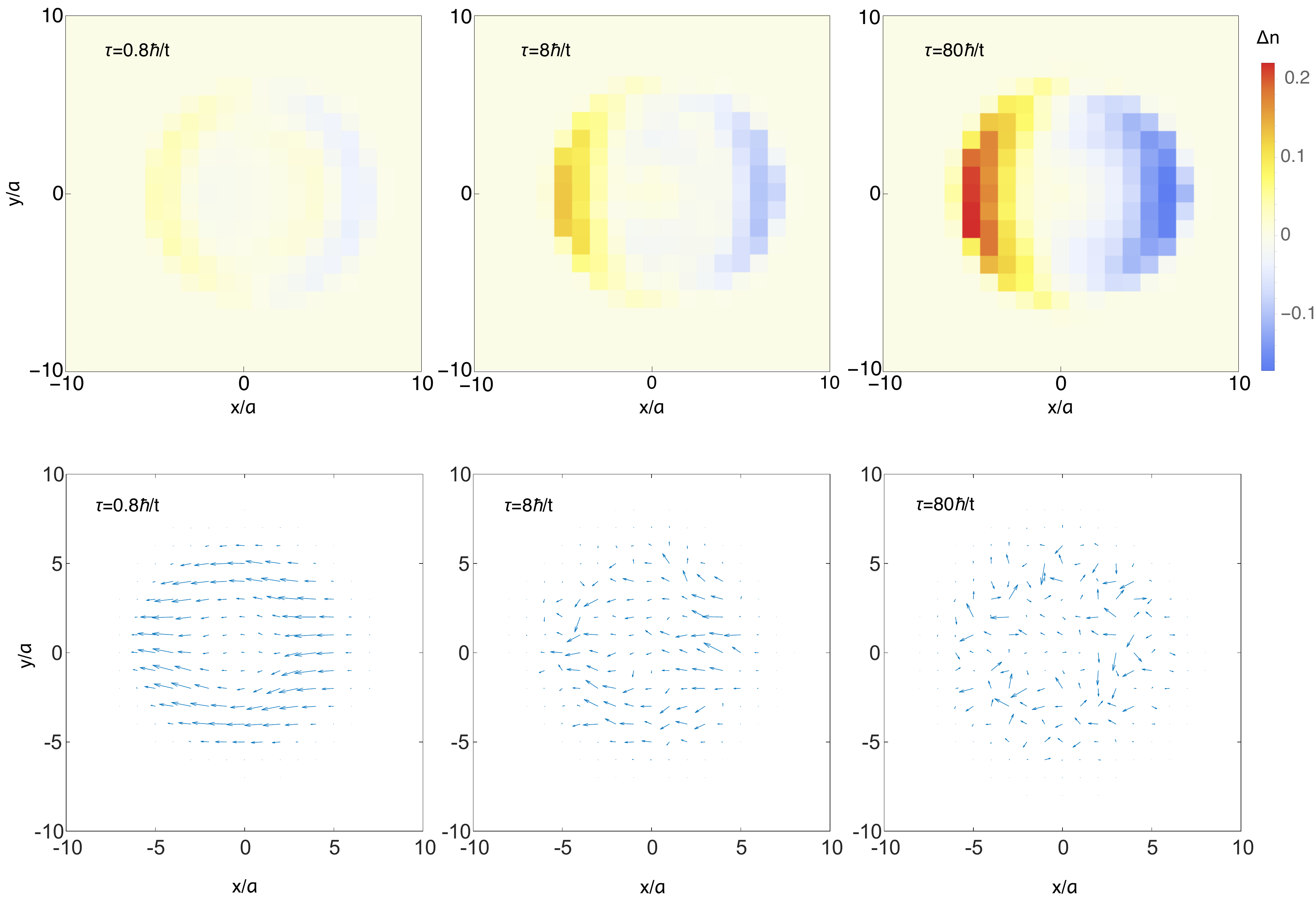}
  \caption{(Color online)Top panel: the same as Fig.~\ref{fig_density_U_8} but for an interaction strength of $U=26t$ (parameters of Fig.~\ref{fig_com_time_8_48}d) which corresponds to a Bose-glass in the center of the trap (central arrow in Fig.~\ref{fig_phasediagram}). Bottom panel: the disorder averaged local current flow with the same parameters in the top panel. At $\tau=0.8\hbar/t$ we see the first mechanism of transport where at short times the edge state quickly slips around the central insulator, depicted in the left panel of Fig.~\ref{fig_schematic_flow}.  At $\tau=8\hbar/t$ and $80\hbar/t$ we see a different, two-step mechanism of transport, depicted in the right panel of Fig.~\ref{fig_schematic_flow}.  Here the particles first slowly tunnel out of the central insulator to, in the second step, get pushed quickly along the edges.  In the two right snapshots, lateral tunneling events and a shrinking of the central insulator can be seen.}
  \label{current_flow_U_26}
\end{figure*}
\subsection{Superfluid Dynamics}
\label{sec_superfluidcm}

We start with the center of mass dynamics of the superfluid (see the right vertical arrow in Fig.~\ref{fig_phasediagram}). Figs.~\ref{fig_com_time_8_48}a-c show damped oscillations of the superfluid in the presence of disorder.  For large interaction strengths the superfluid quickly relaxes to sit at the trap center, Fig.~\ref{fig_com_time_8_48}c.   Here the interplay of interactions and disorder dephase the initial state to effectively relax the system to the ground state of the new trap position.  This is in contrast to the motion of a non-interacting condensate in a trap where  revival occurs at the time scale of inverse trap frequency. \cite{rey:2005, rigol:2005}.  Fig.~\ref{fig_density_U_8} plots the local density at three different times to demonstrate the center of mass oscillations of the superfluid. 

\subsection{Short-Time Insulator Dynamics}
\label{sec_insulatorlong}

\begin{figure}[t]
  \includegraphics[width=0.95\linewidth]{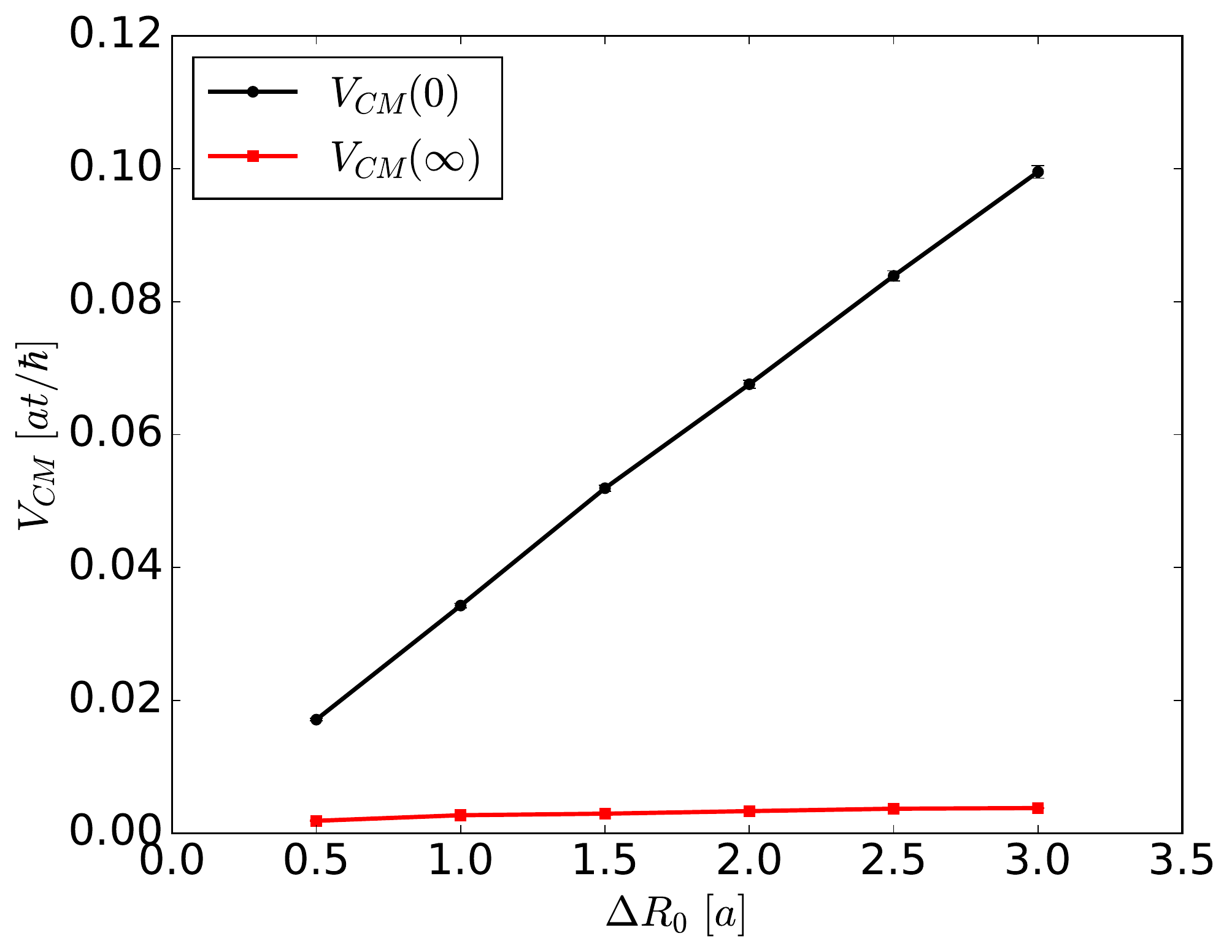}
  \caption{(Color online) The disordered averaged center of mass velocity at short times [Eq.~(\ref{eq_vcm_short})] and long times [Eq.~(\ref{eq_vcm_long})] plotted as a function of the trap shift for an interaction strength of $U=26t$. For short times, the center of mass velocity responds linearly to the length of the trap shift. For long times, the center of mass velocity barely change with the trap shift. }
  \label{fig_velocity_different_shifts}
\end{figure}

We now turn to the center of mass trap dynamics in regimes where either the Bose-glass or Mott insulator lies in the center of the trap (see the left two vertical arrows in Fig.~\ref{fig_phasediagram}).  The state propagated in Fig.~\ref{fig_com_time_8_48}d has a Bose-glass in the center while Fig.~\ref{fig_com_time_8_48}f has a Mott insulator.  Fig.~\ref{fig_com_time_8_48}e shows an intermediate regime on the border of the Mott/Bose-glass transition.  On short time scales, the Bose-glass and Mott states should display insulating behavior because they are both locally gapped.

Figs.~\ref{fig_com_time_8_48}d-f show two distinct regimes: short times ($\tau \lesssim \hbar/t$) and long times ($\tau \gg \hbar/t$).  At short times the center of mass quickly slips to a new value.  Here the initial slip is most prominent in the Bose-glass regime, Fig.~\ref{fig_com_time_8_48}d.  The next section will discuss the long-time dynamics.

To study the motion of the particles with an insulator in the center, we compute
\begin{eqnarray}
I_{j}^{\alpha}=-\mathrm{i}\left( \langle b_j \rangle^{\ast} \langle b_{j+\hat{e}_{\alpha}} \rangle  - \langle b_{j+\hat{e}_{\alpha}} \rangle^{\ast} \langle b_{j} \rangle \right),
\label{eq_current}
\end{eqnarray}
 which is the local current \cite{pethick:2008} flowing from the site ${j}$ to the site ${j}+{\hat{e}}_\alpha$. By plotting the local current distribution in the trap, we can see how the particles move at certain times.
The left panels of Fig.~\ref{current_flow_U_26} shows the density and local current for $U=26 t$ (the Bose-glass regime) after a very short time, respectively.  Here we see that at short times the density flows in a ring around the edge.  

To study the response of the center of mass velocity under an applied field at short times, we compute the initial center of mass velocity 
\begin{eqnarray} 
V_{\text{CM}} (0)\equiv\lim_{\tau\rightarrow 0}\frac{1}{\tau} \int_{0}^{\tau}d\tau'V_{\text{CM}} (\tau').
\label{eq_vcm_short}
\end{eqnarray} 
In the numerical calculation, we choose $\tau=0.2\hbar/t$ to compute the initial velocity. Fig.~\ref{fig_velocity_different_shifts} shows that the initial center of mass velocity responds linearly to the applied force.

\begin{figure}[t]
  \includegraphics[width=0.7\linewidth]{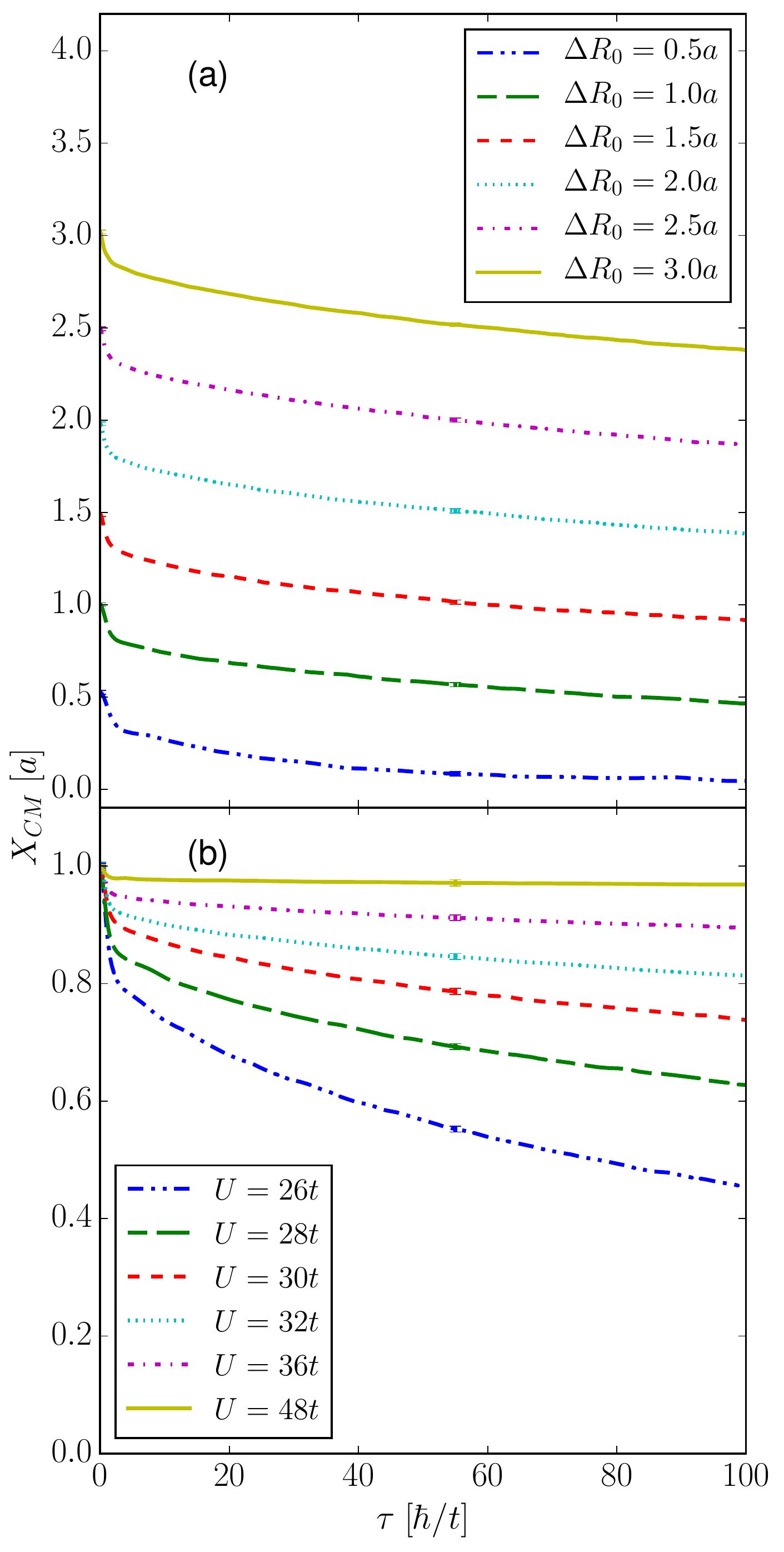}
  \caption{(Color online) The disorder averaged center of mass position plotted as a function of time for: (a) several different trap shifts for $U=26$ and (b) several different $U$ for $\Delta R_{0}=a$.  The model parameters are otherwise the same as Fig.~\ref{fig_density_radius}.  In panel (a) we see that different trap shifts do not change the long time center of mass velocity (the slope is nearly constant).  But in panel (b) we see that the slope depends on $U$.  }
  \label{fig_com_time_BG_MI_variousD}
\end{figure}

The redistribution of the initial edge density corresponds to motion of the edge superfluid at velocities consistent with the band motion of individual edge particles.  The left panel of Fig.~\ref{fig_schematic_flow} depicts short-time edge flow.  We verify that the times required for edge particles to move a distance on the order of the cloud size are consistent with our simulations.  If we consider a one-dimensional approximation to the edge state, the tight-binding energy of a single particle is $-2t\cos(ka)$ where $k$ is the lattice wavevector. The semi-classical equation of motion under a applied field is $\hbar \mathrm{d}k/\mathrm{d}\tau=\Delta R_0 \Omega$.  For short times, the velocity of the particle is then $2t\Delta R_0 \Omega \tau/\hbar^2$, which is linear in the applied field. The $U=26t$ insulator extends 4 lattice sites in the trap with trapping potential $\Omega=0.02U$.  The total time for the edge particle to travel the length of the system under the trap shift $\Delta R_0=a$ is then $2.77 \hbar/t$, consistent with the result from our numerical simulations, $\approx2.5 \hbar/t$.

Our results show that the initial dynamics of trapped insulators reflect the motion of edge states.  The edge moves quickly, on time scales consistent with the motion of free particles. For short times, the center of mass velocity responds linearly to an applied potential.  We now turn to the long-time regime.

\subsection{Long-Time Insulator Dynamics}
\label{sec_insulatorlong}

After the initial slip we find a regime of \emph{constant center of mass velocity} at long times, $\tau \gg \hbar/t$ (Figs.~\ref{fig_com_time_8_48}d-f).  A naive expectation would be that the constant velocity is established by a slow drift of the center of mass as a whole where the disorder strength sets a scattering rate.  We find that the this picture is not correct.  Fig.~\ref{fig_com_time_BG_MI_variousD}a shows that the long-time center of mass velocity does not change with the application of a linear potential and is therefore terminal (recall that the trap shift, $\Delta R_{0}$, effectively applies a linear potential).  This behavior contrasts with behavior expected from a Drude-type model of transport where the center of mass as a whole would respond linearly to a weak potential.

To better understand the mechanism behind the terminal velocity at long times we study the density as a function of time.  The density in the trap plotted for several different times (the two right panels in Fig.~\ref{current_flow_U_26}) reveals a two step process for long-time edge flow (right panel of Fig.~\ref{fig_schematic_flow}).  First, a particle tunnels out of the center to the edge at a rate determined by the competition between the local gap in the center and the trapping potential.  In the second stage, the particle quickly moves along the edge to redistribute the center of mass.  Repeated tunneling out of the center followed by flow along the edge then results in a center of mass shift. Slow tunneling into the edge allows edge flow at long times and therefore dictates the time scale for center of mass motion. 

To verify the two-step picture we construct a simple model to estimate the velocity at long times.  Appendix~\ref{sec_two_site} uses a two-site model to show that in the absence of disorder the tunneling rate out of the center only depends on the trapping frequency, tunneling, and interaction strength.  The velocity in our two-site model is therefore independent of the trap shift.  We also find that the velocity expected from our two-site model compares well with our simulations. 

The two-step picture and our simple model do suggest that the center of mass velocity should depend on $U$.  Fig.~\ref{fig_com_time_BG_MI_variousD}b shows that the  long-time center of mass velocity varies with $U$.  Here stronger $U$ suppresses tunneling out of the bulk because the trap must compensate with high edge energies to confine the system.  We have checked that the $U$ dependence of the two-site model is consistent with our simulation.

We now discuss the long-time center of mass velocity as a direct measure of insulating behavior in the trap \cite{scarola:2006a}.  We compute: 
\begin{eqnarray} 
V_{\text{CM}} (\infty)\equiv\lim_{\tau\rightarrow\infty}\frac{1}{\tau} \int_{\tau_{0}}^{\tau}d\tau'V_{\text{CM}} (\tau'), 
\label{eq_vcm_long}
\end{eqnarray} 
where the limit $\tau\rightarrow\infty$ is understood to imply a long time average such that the insulating states do not reach the center of the trap.  We choose $\tau_{0}=20 \hbar/t$ to include only long-times by avoiding integration over the short-time dynamics.  We expect that $V_{\text{CM}} (\infty)$ should vanish for both immobile states and for states that oscillate about the trap center because there is no net center of mass velocity for an oscillating state.

\begin{figure}[t]
  \includegraphics[width=0.95\linewidth]{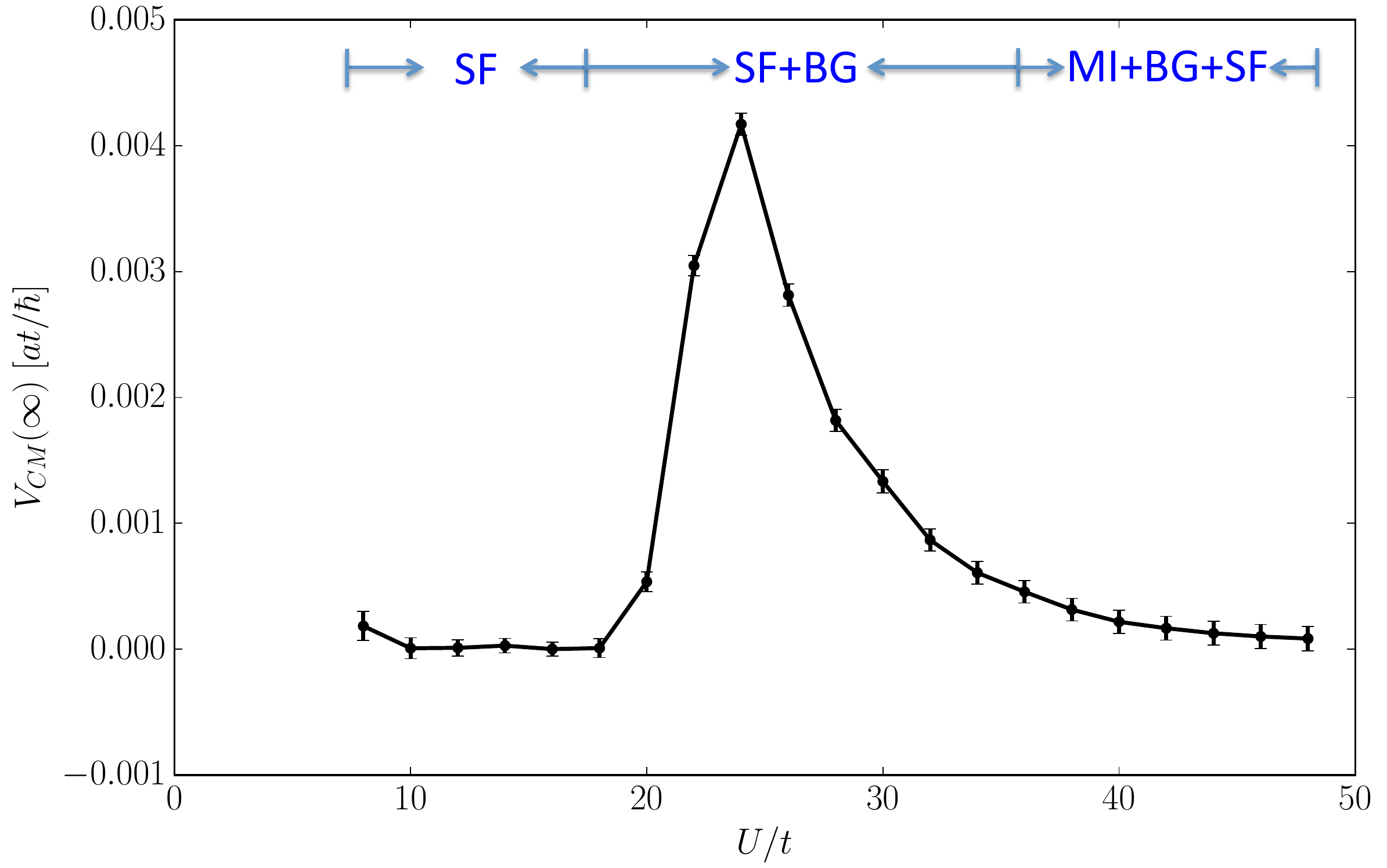}
  \caption{(Color online) The disorder averaged center of mass velocity at long times [Eq.~(\ref{eq_vcm_long})] plotted as a function of interaction strength.  For weak interactions the entire system is in the superfluid (SF) phase and frequent center of mass oscillations results in a zero net velocity.  For intermediate interaction strengths the Bose-glass (BG) lies in the trap center.  Here the two-step edge flow mechanism depicted in the right panel of Fig.~\ref{fig_schematic_flow} allows a small but finite center of mass velocity.  At large interaction strengths the center of the system is in the Mott regime (MI) where tunneling into the edge (and therefore the center of mass velocity) is strongly suppressed.  }
  \label{fig_velocity_long_vs_U}
\end{figure}

Fig.~\ref{fig_velocity_long_vs_U} plots the long-time center of mass velocity in the trap as a function of $U$.  Varying $U$ allows access to the superfluid, Bose-glass, and Mott.  We find that $V_{\text{CM}} (\infty)$ is zero in the superfluid phase because the superfluid oscillates in the trap.  But at long times the center of mass performs a slow drift in insulating phases to lead to a non-zero $V_{\text{CM}} (\infty)$.  The Bose-glass phase shows the largest velocity because the tunneling out of the bulk into the edge is the fastest here.  But the large $U$ Mott limit restricts tunneling into the edge to eventually stop center of mass motion.

The long time center of mass velocity responds to disorder and trapping.  We have checked that the center of mass velocity in the insulating phases is suppressed by increasing disorder or trapping frequency.  Here strong disorder and trapping impact both processes depicted in the right panel of Fig.~\ref{fig_schematic_flow}.  They suppress tunneling out of the center into the edge while also restricting flow along the edge.  

\section{Summary}
\label{sec_summary}

We have studied the zero temperature trap dynamics of states prepared in a disordered Bose-Hubbard model.  We have computed the center of mass velocity of a disordered superfluid, a disordered Mott insulator, and a Bose-glass using Gutzwiller mean field theory.   While the superfluid exhibits damped oscillations for strong interaction strengths, the other two states are essentially motionless insulators at the trap center.  

We find that the center of mass velocity of trapped insulators is dominated by the flow of edge states.  We also find two time scales for the trapped insulators.  At short time scales the edge superfluid flows around the insulating bulk to move the center of mass.  But at longer time scales, tunneling out of the bulk to the edge establishes a slow drift of the center of mass.

Our results can be used to identify insulating states in trapped optical lattices.  We find that, at long times, the edge flow around the Bose-glass can lead to center of mass velocities that are much larger than in the Mott.  We have studied parameter regimes that are similar to what has been explored in experiments.  In Ref.~\cite{pasienski:2010}, for example, experiments were done using a speckle disorder potential applied to $^{87}$Rb atoms in a cubic optical lattice.  Here the Hubbard parameters were: $t/U\sim0.005-0.3$ and the trapping potential keeps the density near 1.4 at the trap center.  The speckle disorder leads, most prominently, to an exponentially distributed onsite disorder potential in $\mu_{i}$ \cite{zhou:2010} which differs from our study.  But finite temperature effects should lead to the most important difference between our study and previous experiments.  Our results have so far excluded finite temperature effects while experiments \cite{pasienski:2010} have been done in the regime  $k_{\text{B}}T\sim 3t $. 

As a next step we will include non-zero temperatures in the initial state.  We expect that for very low temperatures, $k_{\text{B}}T\ll t$, our $T=0$ results will be qualitatively similar.  But for intermediate temperatures, $t \lesssim k_{\text{B}}T <U$, thermally assisted tunneling out of the central insulators and into the edge will significantly enhance the center of mass velocity.  Our results therefore offer a lower bound on the center of mass velocity.

\acknowledgements

VWS acknowledges helpful discussions with B. DeMarco. We acknowledge support from AFOSR under grant FA9550-15-1-0445.

\appendix

\section{Two-site Tunneling Time}
\label{sec_two_site}

\begin{figure}[t]
\vspace{0.2cm}
  \includegraphics[width=0.9\linewidth]{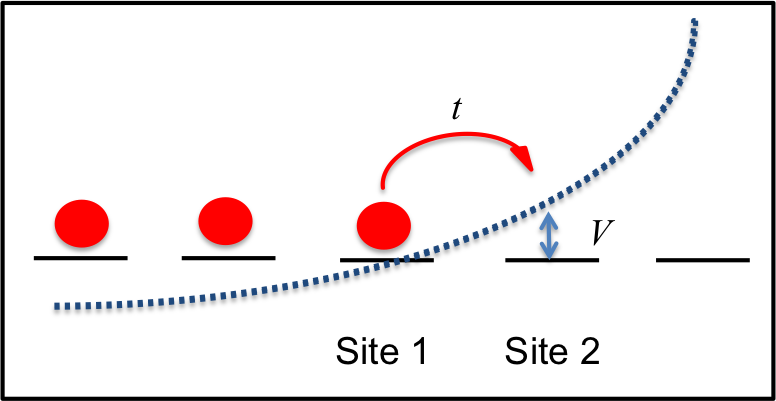}
 \caption{(Color online)  Schematic showing sites of a lattice (dashes) along one direction filled up to site 1 with particles.  Site 2 is an empty edge site.  Tunneling from site 1 to site 2 is controlled with a potential, $V$, at site 2.  The dotted line depicts the parabolic trapping potential which zeroes the edge density. }
 \label{fig_twosites}
\end{figure}

In this section we model the tunneling out of an insulating state at the trap center and into an edge state using a two-site effective model.  Fig.~\ref{fig_twosites} shows part of a Mott insulator along one direction in the lattice.  The curved line depicts the trapping potential which eventually zeroes the edge density.  The trapping potential energy must be on the order of the interaction energy to zero the density.   

We consider a two-site tunneling model Hamiltonian,
\begin{eqnarray}
H_{2} = - t  \left( b_1^\dagger b^{\vphantom{\dagger}}_2 + b_2^\dagger b^{\vphantom{\dagger}}_1 \right) + V   b_2^\dagger b^{\vphantom{\dagger}}_2 ,
\end{eqnarray}
where $V$ is the relative energy gap between the two sites.  $V$ is a fitting parameter on the order of the Mott gap.

Solving the Schr{\"o}dinger equation, we obtain the population on site 2, 
\begin{eqnarray}
n_2 = \frac{2 t^2}{4 t^2+V^2}\left\{ \cos \left[ \sqrt{4 t^2+ V^2} (\tau/\hbar) \right]-1\right\}.
\end{eqnarray} 
The tunneling rate from site 1 to site 2 can then be defined as,
\begin{eqnarray}
\Gamma =\max_{\tau} \frac{dn_2}{d\tau}= \frac{2 t^2}{\hbar\sqrt{4 t^2+ V^2}}.
\end{eqnarray} 
$\Gamma$ gives correct limits:
$\Gamma\rightarrow t$ for $V \rightarrow 0$  and $\Gamma\rightarrow 0$ for $ V \rightarrow \infty$.

\begin{figure}[t]
  \includegraphics[width=0.95\linewidth]{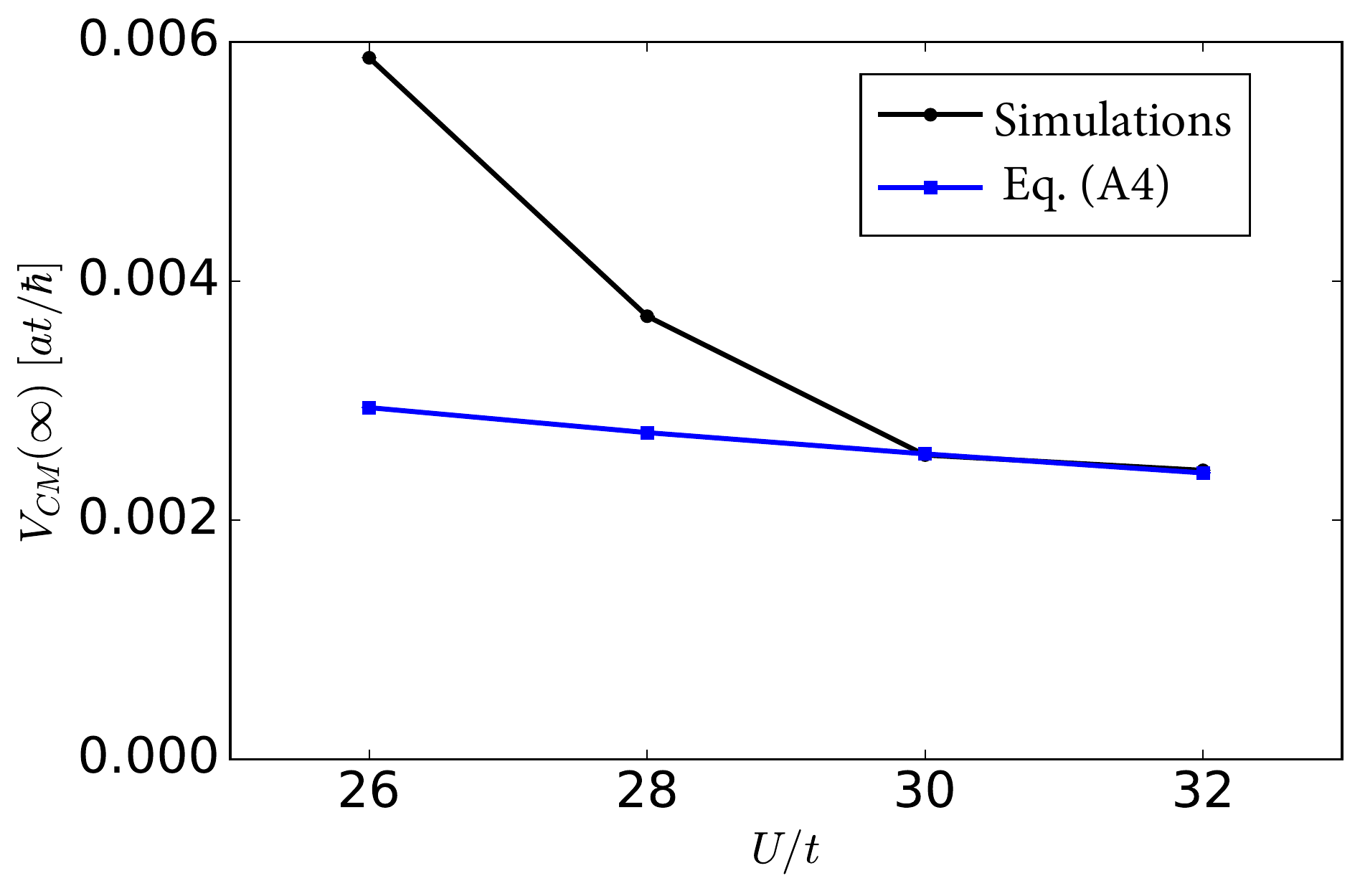}
  \caption{(Color online)  The comparison of the center of mass velocity at long times between the estimate of the two-site model and numerical simulations for various interaction strength.}
  \label{velocity_comparison}
\end{figure}

Using $\Gamma$ as an estimate for the tunneling rate out of the central insulator we can estimate the order of magnitude of the center of mass velocity.  Once the particle tunnels out of the center it then travels along the edge a distance $l$.  The center of mass velocity in this approximation becomes:
\begin{eqnarray}
v= \frac{\Gamma l}{N}=\frac{2 t^2 l}{N\hbar\sqrt{4 t^2+ V^2}},
 \end{eqnarray} 
where $N$ is the total particle number in the trap. To compare this estimate with the our simulations we substitute appropriate parameters. The fitting parameter of the gap $V$ is chosen to be the Mott gap $U$. Fig.~\ref{velocity_comparison} shows that the results obtained from the two-site model are of the same order of magnitude as simulations. 
\bibliography{jabref_database_8_28_15}

\end{document}